\documentclass[aps,prb,twocolumn,superscriptaddress,showpacs]{revtex4}

\usepackage{amsmath,amssymb,latexsym}
\usepackage{graphicx}
\graphicspath{{../fig/}}
\usepackage[dvipdfm,colorlinks=true,linkcolor=blue,citecolor=blue]{hyper ref}
\usepackage{txfonts,mathrsfs}
\usepackage{comment}
\DeclareSymbolFont{symbols}{OMS}{cmsy}{m}{n}
\DeclareSymbolFont{largesymbols}{OMX}{cmex}{m}{n}
\newcommand{\bm}[1]{\boldsymbol #1}

\begin{document}

\title{
Nonequilibrium dynamical cluster theory\\
%a momentum-dependent evolution
}

\author{Naoto Tsuji}
\affiliation{Department of Physics, University of Tokyo, Tokyo 113-0033, Japan}
\author{Peter Barmettler}
\affiliation{D\'epartement de Physique Th\'eorique, Universit\'e de Gen\`eve, 1211 Gen\`eve, Switzerland}
\author{Hideo Aoki}
\affiliation{Department of Physics, University of Tokyo, 113-0033 Tokyo, Japan}
\author{Philipp Werner}
\affiliation{Department of Physics, University of Fribourg, 1700 Fribourg, Switzerland}
%\email[]{}
%\homepage[]{}
%\thanks{}
%\altaffiliation{}

\begin{abstract}

We study the effect of spatially nonlocal correlations on the nonequilibrium dynamics of interacting fermions by 
constructing the nonequilibrium dynamical cluster theory, a cluster generalization of the nonequilibrium dynamical mean-field theory (DMFT).
%which describes a momentum-dependent nonequilibrium relaxation in correlated quantum many-body systems. 
The formalism is applied to interaction quenches in the Hubbard model in one and two dimensions, and the results are  
compared with data from single-site DMFT, the time-dependent density matrix renormalization group, and lattice perturbation theory. 
%We show that the nonlocal correlations play an essential role: 
%In one dimension, the nonlocal correlations enhance non-damped oscillations in double occupancy and momentum distribution, 
%preventing the system form thermalizing.
Both in one and two dimensions the double occupancy quickly thermalizes, while the momentum distribution relaxes only on much longer time scales. 
%For the square lattice we observe a strong momentum dependence in the dynamics of the momentum distribution near the Fermi energy.
For the two-dimensional square lattice we find a strongly momentum-dependent evolution of the momentum distribution around the Fermi energy, with 
a much faster relaxation near the momenta $(0,\pi)$ and $(\pi,0)$ %in the Brillouin zone
than near $(\pi/2,\pi/2)$.  This result is interpreted as 
reflecting the momentum-anisotropic quasiparticle lifetime of the marginal Fermi liquid.  
The method is further applied to the two-dimensional Hubbard model driven by a dc electric field, where the damping of the Bloch oscillation of the current is found to be less effective than predicted by 
DMFT and lattice perturbation theory.
%We also In one dimension, on the other hand, WHAT DO WE WANT TO SAY ABOUT 1D RESULTS?
\end{abstract}

%\collaboration{}
%\noaffiliation

\date{\today}


\maketitle

\section{Introduction}

Simulating the nonequilibrium dynamics of microscopic models for % Hamiltonian for 
quantum many-body systems is a great computational challenge,\cite{PolkovnikovSenguptaSilvaVengalattore2011}  
but such calculations may provide new insights into the role of strong correlations in high-temperature superconductors and other correlated systems
by disentangling complex fluctuations along the real-time axis.
So far, various approaches have been proposed. One focus has been on 
one-dimensional (1D) systems, for which the time-dependent density matrix renormalization group (DMRG) \cite{Daley2004, WhiteFeiguin2004, Schollwoeck2005}
and its variants have provided accurate results for the real-time evolution.
Another approach comes from the opposite limit of infinite dimensions,\cite{MetznerVollhardt1989}  
where the nonequilibrium dynamical mean-field theory (DMFT),\cite{GeorgesKotliarKrauthRozenberg1996, SchmidtMonien2002, FreericksTurkowskiZlatic2006, Aoki2013} which 
incorporates temporal fluctuations but 
approximates the self-energy as a spatially local function, becomes an exact scheme.
However, the nonequilibrium properties of quantum systems in two dimensions, which lies in between these two extremes, remain 
far from being theoretically understood.
%are yet to be elucidated theoretically.

From an experimental point of view, too, the dynamics of two-dimensional (2D) quantum systems is of particular interest.  Recent  
time- and angle-resolved photoemission spectroscopy (ARPES) experiments start to reveal temporal evolutions of 
the occupation $n(\bm k,\omega,t)$ for correlated electrons in layered compounds.\cite{Perfetti2007,Schmitt2008,Graf2011,Cortes2011,Smallwood2012} 
For example, 
%an ultrafast pump-probe ARPES experiment reveals 
it has been shown that the quasiparticle recombination in the $d$-wave superconductor 
Bi$_2$Sr$_2$CaCu$_2$O$_{8+\delta}$ occurs faster away from the ``nodal line" ($k_x=\pm k_y$) in the Brillouin zone than near the nodal line.\cite{Smallwood2012} 
This kind of momentum-dependent relaxation dynamics can be related to 
%the anisotropic Cooper pairing in momentum space and hence 
{\it nonlocal quantum correlations}, which should become essential in 
low-dimensional quantum systems. We can then pose the following questions:
What role do nonlocal correlations play in low-dimensional correlated systems out of equilibrium? 
And how can we take account of these effects systematically in real-time simulations?

%These have motivated us in this Letter to present 
Motivated by these questions, we present and test here 
a theoretical approach, namely the {\it nonequilibrium dynamical cluster theory}, 
which is the cluster extension of the nonequilibrium DMFT, or the nonequilibrium generalization of dynamical cluster theories.\cite{Maier2005} 
In the DMFT formulation, we map a lattice model to a single-site impurity embedded in a dynamical mean field, 
%which is defined in such a way that the local part of the lattice Green's function is reproduced. 
using the assumption of a local self-energy, thereby neglecting nonlocal dynamical fluctuations.  
In cluster formalisms, this restriction is overcome by mapping the system onto a finite-size cluster problem with a spatially correlated dynamical mean field. 
%that is determined self-consistently. 

%We extend the method to nonequilibrium.  
We then apply this scheme to the interaction-quench problem in the Hubbard model in 1D and 2D,
changing the strength of the interaction abruptly in time. 
%as one of the simplest ways of driving the system out of equilibrium.
In cold atoms, where effective interactions can be tuned using Feshbach resonaces or by changing the depth of optical lattice potentials, quantum
quenches have become a standard procedure to trigger nonequilibrium dynamics,\cite{Greiner2002,Chen2011,Trotzky2012,Cheneau2012}  
%Since it can be realized in cold-atom experiments using the Feshbach resonance or modulating the optical lattice potential,
and the problem has attracted broad theoretical interests.\cite{KollathLauchliAltman2007, ManmanaWesselNoackMuramatsu2007, MoeckelKehrein2008, EcksteinKollarWerner2009, 
BarmettlerPunkGritsevDemlerAltman2009, Karrasch2012, WernerTsujiEckstein2012, Hamerla2013a, Tsuji2013, Hamerla2013b} 
A naive expectation is that after the quench the system is highly excited and is characterized by a high effective temperature, so that
the nonlocal correlations might be wiped out, as in equilibrium at high temperatures. However, we will show that
%as the dynamical cluster theory shows,
in 2D the momentum distribution, after experiencing prethermalization,\cite{BergesBorsanyiWetterich2004, MoeckelKehrein2008, EcksteinKollarWerner2009}  
%starts to relax differently in each momentum sector: 
exhibits a {\it momentum-dependent relaxation dynamics}: 
the distribution relaxes to the thermal one 
faster in the antinodal region [around $(0,\pi)$ or $(\pi,0)$ in the Brillouin zone] than in the nodal region 
[around $(\pi/2,\pi/2)$]. The momentum-dependent relaxation, observable only when 
we go from the single-site to the cluster formalism, comes from the nonlocal correlations.  
Our finding is consistent with the quasiparticle lifetime of the marginal Fermi liquid, which is highly anisotropic in momentum space.  
%A peculiar form of the self-energy as a function of energy is consistent with a marginal Fermi liquid. 

We also examine a 1D system, 
where we can benchmark the cluster calculations rigorously by comparing the time-evolution to numerically exact DMRG results,
and test the convergence of the results with respect to the cluster size. 
Here, we find a rapid thermalization of the double occupancy
similar to the 2D case, apart from additional oscillations due to divergences in the density of states at the band edges.  
%@We have also examined the convergence against the cluster size.
%On the other hand, in 1D   HOW DO WE SAY HERE? PB COME BACK

\section{General formulation}

We will formulate the nonequilibrium dynamical cluster theory by taking the Hubbard model as an example.
The time-dependent Hamiltonian is given by 
\begin{align}
H(t)
&=
-J\sum_{\langle ij \rangle, \sigma} (c_{i\sigma}^\dagger c_{j\sigma}+\mbox{H.c.})
-\mu \sum_{i,\sigma} \hat{n}_{i\sigma}
%\nonumber
%\\
%&\quad
+U(t)\sum_i \hat{n}_{i\uparrow}\hat{n}_{i\downarrow},
\nonumber
\end{align}
where $c_{i\sigma}^\dagger$ creates a lattice fermion at the $i$th site with spin $\sigma$,
$\hat{n}_{i\sigma}\equiv c_{i\sigma}^\dagger c_{i\sigma}$,
$J$ is the hopping amplitude, $\mu$ the chemical potential, $U$ the (time-dependent) interaction strength,
and the sum $\langle ij \rangle$ is taken over nearest-neighbor sites. 
There are two well-established constructions for the cluster mapping: the cellular DMFT
\cite{Lichtenstein2000,Kotliar2001} and the dynamical cluster approximation (DCA).\cite{Hettler1998,Hettler2000}
Here we adopt the DCA, since it preserves the periodicity of the lattice structure by construction.
This enables us to use the diagonal (momentum) representation for the cluster Green's function, while in the cellular DMFT the cluster Green's function has to be represented in real space with off-diagonal elements.
The cluster Dyson equation, which we shall introduce below, then becomes a ``matrix'' integral-differential equation, 
which is hard to solve for large size clusters [with the computational cost scaling as $O(N_c^3)$ for clusters of size $N_c$].
This is why we have here opted for the DCA.
%As we shall see, the generalization of DCA to nonequilibrium is straightforward.?

The cluster reference system is defined by the action 
\begin{align*}
\mathcal S_{\rm clust}[\Delta]
&=
-J\int_{\mathcal C} dt \sum_{\langle ij \rangle, \sigma} d_{\bm R_i \sigma}^\dagger(t) d_{\bm R_j \sigma}(t)
-\mu\int_{\mathcal C} dt \sum_{i\sigma} \hat{n}_{\bm R_i\sigma}(t)
\\
&\quad
+\int_{\mathcal C} dt U(t)\sum_{i} \hat{n}_{\bm R_i\uparrow}(t) \hat{n}_{\bm R_i\downarrow}(t)
\\
&\quad
+\int_{\mathcal C} dt \int_{\mathcal C} dt' \sum_{ij\sigma}
d_{\bm R_i\sigma}^\dagger(t)\Delta_{\sigma}(\bm R_i-\bm R_j;t,t')d_{\bm R_j\sigma}(t'),
\end{align*}
%[SHOULD ADD THE CLUSTER HOPPING MATRIX, IF WRITTEN IN TERMS OF THE HYBRIDIZATION FUNCTION]
where $d_{\bm R_i\sigma}^\dagger$ creates a cluster fermion at a cluster site $\bm R_i$,
$\Delta_\sigma(\bm R;t,t')$ is the hybridization function that will be determined self-consistently,
$\hat{n}_{\bm R\sigma}=d_{\bm R\sigma}^\dagger d_{\bm R\sigma}$, and the time integral is taken along 
the Kadanoff-Baym contour $\mathcal C$,\cite{Rammer} running along $t=0 \to t_{\rm max} \to 0 \to -i\beta$ 
(where $t_{\rm max}$ is the maximum time up to which the system is evolved, and $\beta$ the inverse temperature of the initial thermal state). 
With this action, we define the cluster Green's function
$G_\sigma^{\rm clust}(\bm R-\bm R';t,t')=-i\langle \mathcal T_{\mathcal C} d_{\bm R\sigma}(t)d_{\bm R'\sigma}^\dagger(t')\rangle_{\mathcal S_{\rm clust}}$
with $\mathcal T_{\mathcal C}$ the contour-ordering operator along $\mathcal C$
and $\langle \cdots \rangle_{\mathcal S_{\rm clust}}={\rm Tr}(\mathcal T_{\mathcal C} e^{-i\mathcal S_{\rm clust}}\cdots)/{\rm Tr}(\mathcal T_{\mathcal C} e^{-i\mathcal S_{\rm clust}})$.

%Let $\bm K$ be reciprocal wave vectors of $\bm R$. We can fourier transform the cluster Green's function,
If we denote by $\bm K$ the wave vector reciprocal to $\bm R$, we can write the Fourier-transformed cluster Green's function as
\begin{align}
G_\sigma^{\rm clust}(\bm K;t,t')&=\sum_{j} e^{-i\bm K\cdot\bm R_j}G_\sigma^{\rm clust}(\bm R_j;t,t').
\nonumber
\end{align}
The Brillouin zone is divided into $N_c$ sectors, each of which is centered at the corresponding $\bm K$.
There are various choices of clusters. We adopt two cases, 
\begin{align*}
&\mbox{A: } \bm K_{x,y}=2n_{x,y}\pi/N_c, \\
&\mbox{B: } \bm K_{x,y}=(2n_{x,y}-1)\pi/N_c
\end{align*}
($n_{x,y}$: integers).
%\textcolor{red}{\sout{and average the final results.}}
In the lattice problem, an arbitrary wave vector $\bm k$ can be written as $\bm K+\tilde{\bm k}$, where $\tilde{\bm k}$ represents the relative 
momentum from the center of the momentum sector.
The mapping from the lattice to the cluster problem (i.e., the choice of the hybridization function $\Delta_\sigma$) is defined 
such that the cluster Green's function is reproduced by the lattice Green's function averaged over the corresponding momentum sector,
\begin{align}
G^{\rm clust}_\sigma[\Delta](\bm K;t,t')
&=
\frac{N_c}{N}\sum_{\tilde{\bm k}} G^{\rm lat}_\sigma(\bm K+\tilde{\bm k};t,t'),
\nonumber
\end{align}
with $N$ the total number of $k$ points. 
%Each Green's function is related to the self-energies via the cluster Dyson equation
The Green's functions and self-energies are related via the cluster Dyson equation,
\begin{align}
&(i\partial_t+\mu)G_\sigma^{\rm clust}(\bm K)-\Delta_\sigma(\bm K)\ast G_\sigma^{\rm clust}(\bm K)
-\Sigma_\sigma^{\rm clust}(\bm K)\ast G_\sigma^{\rm clust}(\bm K)
\nonumber
\\
&\quad
=\delta_{\mathcal C}(t,t'),
\nonumber
\end{align}
with $\ast$ representing a convolution along the contour $\mathcal C$, 
and the lattice Dyson equation,
\begin{align}
&(i\partial_t+\mu)G_\sigma^{\rm lat}(\bm k)-\epsilon(\bm k)\ast G_\sigma^{\rm lat}(\bm k)
-\Sigma_\sigma^{\rm lat}(\bm k)\ast G_\sigma^{\rm lat}(\bm k)
%\nonumber \\&\quad
=\delta_{\mathcal C}(t,t'),
\nonumber
\end{align}
where $\epsilon(\bm k)=-2J\sum_{i=1}^d \cos k_i$ is the band dispersion,
and $\delta_{\mathcal C}(t,t')$ the contour delta function defined on $\mathcal C$.
In the nonequilibrium DCA, we identify the lattice self-energy with the cluster self-energy,
\begin{align}
\Sigma^{\rm lat}_\sigma(\bm K+\tilde{\bm k};t,t')
&=
\Sigma^{\rm clust}_\sigma(\bm K;t,t'),
\nonumber
\end{align}
that is, we neglect the $\tilde{\bm k}$ dependence of $\Sigma_\sigma^{\rm lat}(\bm K+\tilde{\bm k};t,t')$.
In this way, the problem is reduced to solving the cluster problem, for which
one may use several possible solvers developed for the nonequilibrium DMFT, 
e.g., the weak-coupling perturbation theory,\cite{EcksteinKollarWerner2010, Tsuji2013b} quantum Monte Carlo,\cite{WernerOkaMillis2009}
the noncrossing approximation (NCA),\cite{Eckstein2010} and exact-diagonalization-based approaches.\cite{Arrigoni2013, Gramsch2013} 

In the present formalism, spatial correlations are systematically included within a finite range cutoff $L\sim N_c^{1/d}$.
In the large cluster-size limit ($N_c\to\infty$), the formalism should become exact in arbitrary dimensions.
A virtue of the nonequilibrium DCA is that it provides a self-consistency scheme that updates 
the ``noninteracting part'' ($\Delta_\sigma$) of the action, so that it can describe ``thermalization'' in the long-time limit.
This discriminates it from other existing approaches that capture nonlocal correlations in the time evolution.
For instance, the conventional lattice perturbation technique expands the self-energy in terms of 
the noninteracting lattice Green's function, so that the memory of the initial state is kept permanently.
In cluster perturbation theory,\cite{Balzer2011,Knap2011,Balzer2012} which decomposes the system into clusters 
and treats the inter-cluster connections perturbatively, the feedback to the exactly solved subsystems is limited.  
The dual-fermion approach provides another path to extend the DMFT, but 
its application to nonequilibrium situations is so far limited to a small cluster and impurity problem.\cite{Jung2012,Munoz2013} 
Very recently, the equation-of-motion method has been applied to the 2D Hubbard model.\cite{Hamerla2013b}
This scheme allows to compute numerically exact results, but only up to relatively short times.  

If we concentrate on the weak-coupling regime at half-filling, 
we can employ the iterative perturbation theory (IPT) as a cluster solver: 
\begin{align}
\Sigma_{\sigma}^{\rm clust}(\bm R;t,t')
&=
U(t)U(t')\mathcal G_{0\sigma}(\bm R;t,t')\mathcal G_{0\bar{\sigma}}(-\bm R;t',t)
\mathcal G_{0\bar{\sigma}}(\bm R;t,t').
\nonumber
\end{align}
Here $\mathcal G_{0\sigma}(\bm R;t,t')$ is the cluster Weiss mean-field propagator defined by
\begin{align}
&(i\partial_t+\mu)\mathcal G_{0\sigma}(\bm K)-\Delta_\sigma(\bm K)\ast \mathcal G_{0\sigma}(\bm K)
=\delta_{\mathcal C}(t,t').
\nonumber
\end{align}
We note that IPT as an impurity solver in nonequilibrium DMFT calculations works adequately for $U$ smaller than or equal to half the bandwidth.\cite{Tsuji2013b} %We also use IPT for the nonequilibrium DMFT solver in the following.

\begin{figure}[t]
\includegraphics[width=8cm]{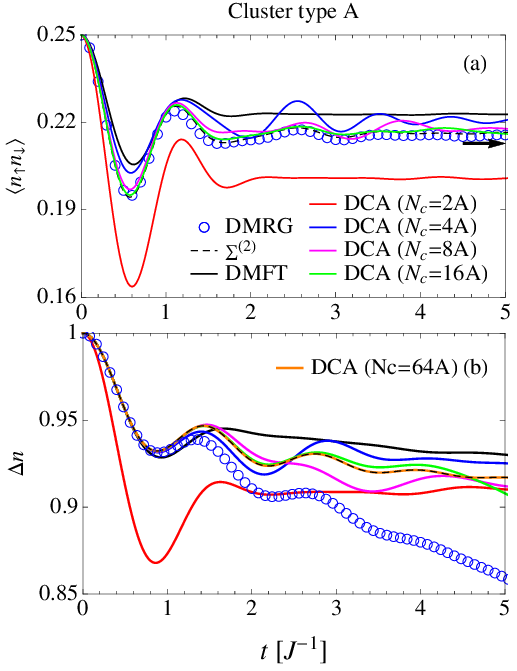}
\caption{(a) The double occupancy and (b) the jump in the momentum distribution for a quench $U/J=0\to 1$ in the 1D Hubbard model
calculated by DCA with cluster type A, and compared with other methods.
The arrow in (a) indicates the thermal value of the double occupancy evaluated from the finite-temperature DMRG.}
\label{1d hubbard-A}
\end{figure}
%We first apply the DCA method to the one dimensional system. 

We can use the time-dependent DMRG \cite{Daley2004,WhiteFeiguin2004,Schollwoeck2005} to benchmark the DCA result for the 1D system. By using a matrix-product state formalism in the thermodynamic limit,\cite{Vidal2007} we can get rid of finite-size effects, with the accuracy of the results only limited by the number of DMRG states, which is chosen here to be $D=800$ for the initial state and up to $D=3600$ for the subsequent time evolution. The maximum truncation of the density-matrix eigenvalues is $\epsilon=10^{-7}$, leading to numerical errors much smaller than the symbol sizes in the figures. The initial state is generated by an imaginary-time evolution with an explicit orthogonalization scheme applied.\cite{McCulloch2008} We have also performed finite-temperature DMRG \cite{Verstraete2004,Zwolak2004,Feiguin2005} calculations to  compare the long-time properties with thermal-equilibrium results.

\begin{figure}[t]
\includegraphics[width=8cm]{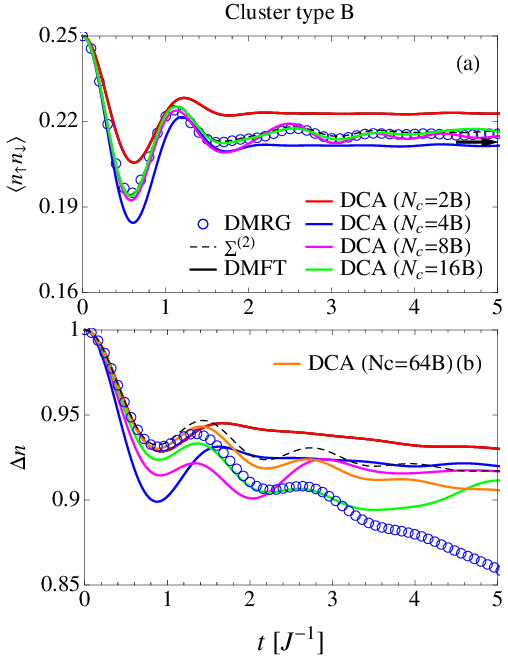}
\caption{(a) The double occupancy and (b) the jump in the momentum distribution for a quench $U/J=0\to 1$ in the 1D Hubbard model 
calculated by DCA with cluster type B, and compared with other methods.
The arrow in (a) indicates the thermal value of the double occupancy evaluated from the finite-temperature DMRG.}
\label{1d hubbard-B}
\end{figure}

\section{Interaction quench: 1D Hubbard model}
\begin{figure}[t]
\includegraphics[width=8.5cm]{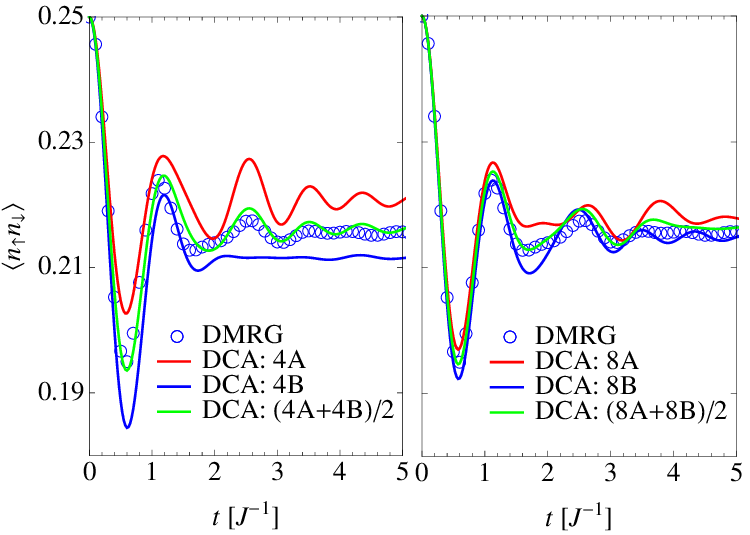}
\caption{
Averaging of the double occupancy obtained by DCA with cluster types A (Fig.~\ref{1d hubbard-A}) and B (Fig.~\ref{1d hubbard-B})
for $N_c=4$ (left) and $N_c=8$ (right).
}
\label{1d hubbard-average}
\end{figure}

As a first application, we study the interaction quench $U(t)=0\to U>0$ for the Hubbard model, starting from the noninteracting 
zero-temperature state.
We plot the time evolution of the double occupancy
$d(t)=\langle \hat{n}_{\uparrow}(t)\hat{n}_{\downarrow}(t)\rangle$, 
along with the jump $\Delta n(t)$ in the momentum distribution 
$n(\bm k,t)=\langle c_{\bm k\sigma}^\dagger(t) c_{\bm k\sigma}(t)\rangle$ at the Fermi energy 
[$\epsilon(\bm k)=\epsilon_F$].
Figure~\ref{1d hubbard-A} (Fig.~\ref{1d hubbard-B}) shows results for the 1D Hubbard model calculated by the DCA with cluster type A (B).
We also plot for comparison the results of DMRG, DMFT, and 
the second-order lattice perturbation theory ($\Sigma^{(2)}$). 
The number of $k$ points is $N=1024$ for the methods other than DMRG. 
Compared to the infinite-coordination Bethe lattice, where $d(t)$
relaxes rapidly,\cite{EcksteinKollarWerner2009} DMFT predicts a damped oscillation in $d(t)$ for the 1D lattice.
DCA with $N_c={\rm 2B}$ (i.e., $N_c=2$ with cluster type B)
%@you have to define 2B etc (something like "Nc=2 in the type B choice of the k points") 
is exactly equivalent to DMFT due to the Brillouin-zone symmetry [Fig.~\ref{1d hubbard-B}(a)],
while the DCA result  for $d(t)$ with $N_c={\rm 2A}$ is overall smaller than that of DMFT having a damped oscillation [Fig.~\ref{1d hubbard-A}(a)].
As we proceed to DCA with $N_c\ge 4$ where the momentum space near the Fermi energy and the band edge can be distinguished, 
the oscillations become more pronounced. This suggests that the oscillation originates from the divergence of the density of states at the band edges in 1D. 
In fact, the oscillation period is roughly $2\pi/(4J)$ (with $4J$ being the bandwidth of the 1D lattice).
As $N_c$ is increased, the cluster-type (A or B) 
dependence becomes weaker, and 
the DCA results converge to the exact DMRG. If we take an average of the results obtained from DCA over the cluster types A and B (Fig.~\ref{1d hubbard-average}), 
the convergence to DMRG in the limit of $N_c\to\infty$ is notably accelerated, 
with a fair agreement already seen even at $N_c=4$.
%the DCA results quickly approach the exact DMRG (a fair agreement is already seen at $N_c=4$).
%[BUT THIS IS ONLY THE CASE IF ONE AVERAGES A AND B. DO WE NEED AN ADDITIONAL ILLUSTRATION FOR THIS?]

Unlike the double occupancy, the jump in the momentum 
distribution $\Delta n$ in DCA [Fig.~\ref{1d hubbard-A}(b) for cluster type A and Fig.~\ref{1d hubbard-B}(b) for type B] 
does not converge rapidly with $N_c$.  
This can be related to the nonlocal nature of the quantity $\Delta n$, which is derived via Fourier transformation  
from the real-space correlation $\langle c_{i\sigma}^\dagger c_{j\sigma} \rangle$.
If we increase $N_c$ up to 64, 
%revision
we observe that DCA+IPT approaches $\Sigma^{(2)}$ for the 1D case.
The deviation from DMRG must be attributed to quantum corrections from higher-order diagrams neglected in IPT.
According to DMFT, $\Delta n(t)$ exhibits a prethermalization plateau \cite{EcksteinKollarWerner2009} 
after a rapid initial drop, which is a characteristic feature of prethermalization.\cite{MoeckelKehrein2008, KollarWolfEckstein2011} 
According to DCA and DMRG, however, a clear prethermalization plateau is not observed.
Instead, similar to $d(t)$, we see an oscillation in $\Delta n(t)$ which does not damp fast, unlike in the higher-dimensional cases. 
The momentum distribution relaxes much more slowly than $d(t)$, and is still far from the thermal distribution with $\Delta n=0$
on the computationally accessible time scale.

While one might think that the 1D Hubbard model, being integrable,\cite{LiebWu1968} should be prevented from thermalization, 
we find that the double occupancy takes, fairly soon after the quench, a value close to the thermal value [arrows in Figs.~\ref{1d hubbard-A}(a) and \ref{1d hubbard-B}(a)].
For integrable models, nonequilibrium states are often described by the generalized Gibbs ensemble \cite{Rigol2007} in terms of a macroscopic number of integrals of motion.
%In certain situations, only a few conserved quantities are sufficient to estimate expectation values of observables \cite{Barmettler2013}.
Our analysis suggests that $d(t)$ is not very sensitive to the nontrivial conserved quantities of the Hubbard model,\cite{Shastry1986} 
and that the total energy along with the total number of particles almost fully describes the stationary value of $d(t)$.  
% and also $\Delta n(t)$ seems to relax to zero even though
%This may be natural since the 1D Hubbard model is integrable.

%Nevertheless, we find that after the quench the double occupancy takes a value close the thermal value  
%that would be realized if the system fully thermalizes, even though the momentum distribution is still far 
%PSEUDOTHERMALIZATION?
%PLEASE ADD MORE DISCUSSION ON 1D RESULTS HERE.

\begin{figure}[t]
\includegraphics[width=8cm]{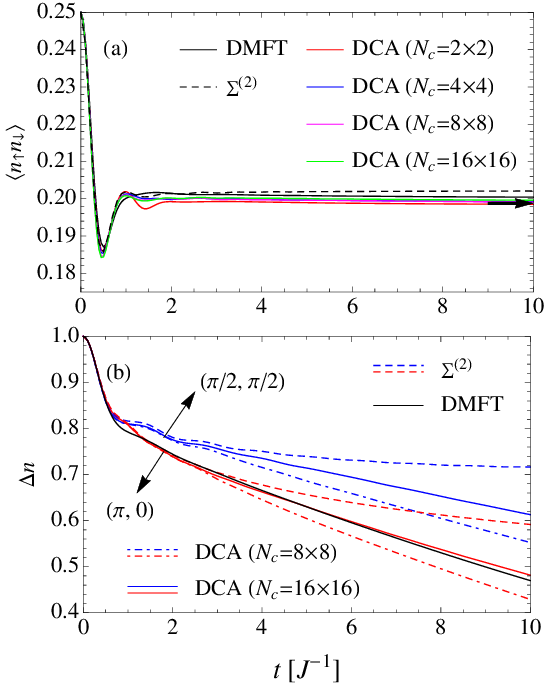}
\caption{(a) The double occupancy and (b) the jump in the momentum distribution for a quench $U/J=0\to 2$ in the 2D Hubbard model, 
obtained by DCA with the average of cluster types A and B, as compared with other methods.
The arrow in (a) indicates the thermal value evaluated from DCA with $N_c=16\times 16$.}
\label{2d hubbard}
\end{figure}

\section{Interaction quench - 2D Hubbard model}
Now let us turn to the interaction quench in the 2D Hubbard model, 
and investigate whether there is a qualitative change in the relaxation dynamics 
when going from 1D to 2D. 
%Moreover the 2D Hubbard model is not integrable. % where the singularity is at the edges of the band.
%Since the 2D model is not integrable, we expect 
%a behavior different from the 1D model. 
In Fig.~\ref{2d hubbard}, we plot the time evolution for $d(t)$ and $\Delta n(t)$ for 2D.
Since the cluster types A and B give quantitatively similar results, 
we take the average of cluster types A and B in the results of DCA to accelerate the convergence with respect to $N_c$, as we did for 1D.
We can see that $d(t)$ quickly relaxes to the thermal value [an arrow in Fig.~\ref{2d hubbard}(a)] without
generating long-lived oscillations. This is similar to the infinite-dimensional case. 
For $d(t)$, we find that the dependence on $N_c$ is quite small (with 
DMFT already providing a good estimate), which 
implies that the nonlocal correlations are less relevant for local quantities in 2D. 

However, if we turn to $\Delta n(t)$ which is a nonlocal quantity, 
we immediately notice that $\Delta n(t)$ now dramatically depends 
on the position along the Fermi surface [while DMFT only gives a momentum-independent $\Delta n(t)$]. 
Note that the DCA results do converge with respect to $N_c$ in the short-time regime up to $tJ\lesssim 2$, 
where the momentum dependence already starts to grow.
In the plot we have focused on the nodal $(\pi/2,\pi/2)$ and antinodal $(\pi,0)$ points, and we consider clusters up to $N_c=16\times 16$ (since we need $N_c\ge 4\times 4$ to distinguish the nodal and antinodal sectors). 
As was the case in 1D, $\Delta n(t)$ is sensitive to $N_c$,  
and even with clusters as large as $N_c=16\times 16$ we still have a finite cluster-size effect.  
It seems that DCA+IPT is approaching $\Sigma^{(2)}$ in the large $N_c$ limit (at least for $\Delta n$) in this interaction range. 
%However, it remains yet to be confirmed since it is hard from a point of view of memory storage
%to take the limit $N_c\to\infty$ keeping $N/N_c$ large enough.
At present, going to larger clusters is technically difficult due to memory limitations, since we have to keep $N/N_c$ large enough
(in Fig.~\ref{2d hubbard} we take $N=256\times 256$).

A salient feature in Fig.~\ref{2d hubbard}(b) is that, while 
$\Delta n(t)$ evolves completely uniformly over the momentum space in the early stage ($tJ\le 0.5$), 
it suddenly starts to exhibit a momentum dependence after that period:  The antinodal point $(\pi,0)$ relaxes faster than
the nodal point $(\pi/2,\pi/2)$, where a slowly damped oscillation appears in the time evolution. The latter is reminiscent of
the 1D results. The DCA simulation suggests that the momentum distribution eventually reaches the thermal distribution with $\Delta n=0$.
If one goes to larger $U$, the momentum variation of $\Delta n(t)$ is weakened.

\begin{figure}[t]
\includegraphics[width=8.5cm]{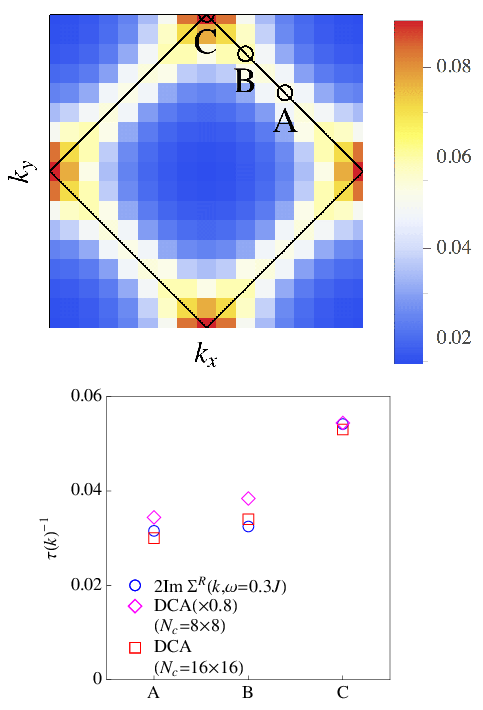}
\caption{(Color online) Top: Color-coded plot of $|{\rm Im}\,\Sigma^R(\bm k,\omega)/\omega|$ at $\omega=0.3J$ for $U/J=2$ and $T=0$.
Solid lines indicate the noninteracting Fermi surface.
Bottom: Inverse of the quasiparticle lifetime $\tau(\bm k)$
estimated with the nonequilibrium DCA, along with that estimated from ${\rm Im}\,\Sigma^R(\bm k,\omega)$ for three positions 
(A, B, and C on the left panel) in the Brillouin zone.}
\label{quasiparticle}
\end{figure}

Now, let us examine what the momentum-dependent relaxation seen in $\Delta n(t)$ implies, based on the quasiparticle picture, which is valid in the 
weak-interaction regime. 
The lifetime $\tau(\bm k)$ of the quasiparticle with energy $\omega$ can be 
evaluated from the equilibrium retarded self-energy,\cite{Hodges1971}
%\cite{AbrikosovGorkovDzyaloshinskiBook},
\begin{align}
\tau(\bm k)^{-1}=2{\rm Im}\Sigma^R(\bm k,\omega).
\nonumber
\end{align}
The 2D Hubbard model on the square lattice at half filling is special, since 
the one-particle dispersion has a van Hove singularity right at the Fermi energy $\epsilon_F$.
This makes the density of states diverging logarithmically, 
and the system behaves as a ``marginal Fermi liquid,'' i.e., ${\rm Im}\,\Sigma^R(\bm k,\omega)\propto \omega$ around $\omega=\epsilon_F$.\cite{Pattnaik1992} 
In the top panel of Fig.~\ref{quasiparticle}, we plot ${\rm Im}\,\Sigma^R(\bm k,\omega)$ obtained from $\Sigma^{(2)}$.
Even in the weak-coupling regime, ${\rm Im}\,\Sigma^R(\bm k,\omega)$ is highly anisotropic in momentum space: It is peaked at $(\pi,0)$, while $(\pi/2,\pi/2)$
exhibits a saddle-point behavior. In the bottom panel of Fig.~\ref{quasiparticle}, we compare the inverse quasiparticle lifetime ${\rm Im}\,\Sigma^R(\bm k,\omega)$
with the relaxation rate $\gamma$ for $\Delta n$. The latter is evaluated by fitting the DCA result for $\Delta n(t)$ [Fig.~\ref{2d hubbard}(b)] with a single exponential $Ae^{-\gamma t}$.
We find that the momentum dependence of $\gamma$ (whose qualitative tendency is independent of $N_c$) 
is well reproduced by ${\rm Im}\,\Sigma^R(\bm k,\omega)$.  
This  suggests that
the momentum-dependent relaxation of $\Delta n(t)$ is in fact 
{\it governed by the quasiparticles}, which have a longer lifetime at $(\pi/2,\pi/2)$.

\section{2D Hubbard model driven by dc fields}

The nonequilibrium DCA proposed here is 
a general framework, which allows to study not only 
nonequilibrium phenomena resulting from interaction quenches but also those induced 
by dc-field quenches.  Let us demonstrate this here for 
the 2D Hubbard model on the square lattice driven by a dc electric field $E$.  
The field is introduced by the Peierls substitution $\epsilon(\bm k)\to \epsilon(\bm k-\bm A(t))$ 
in the noninteracting part of the lattice Hamiltonian with $\bm A(t)=-\bm E t$ the vector potential, where the field is taken to 
be along the diagonal direction, i.e., $\bm E=E(1,1)$.
We switch on the field at $t=0$ with the initial state being 
the noninteracting one at zero temperature. 
The interaction is quenched as $U/J=0\to 2$ at the same time as the field is turned on at $t=0$. A physical observable of interest 
in this situation is the current,
\begin{align*}
j = -i\sum_{\bm k\sigma} v_{\bm k-\bm A(t)}G_{\bm k\sigma}^{{\rm lat},<}(t,t),
\end{align*}
where $v_{\bm k}=\sum_i \partial \epsilon_{\bm k}/\partial k_i$ is the velocity along the $(1,1)$ direction.

\begin{figure}[t]
\includegraphics[width=8cm]{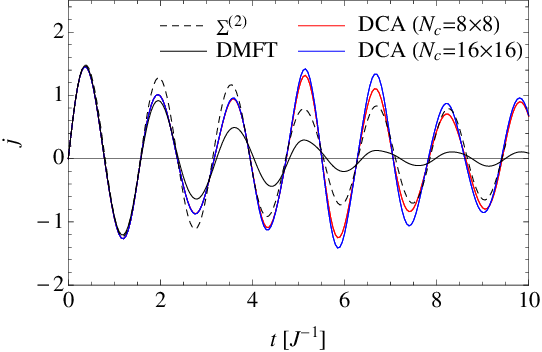}
\caption{
The current in the 2D Hubbard model driven by a dc electric field $E=4$ obtained with DCA 
for $N_c=8\times 8, 16\times 16$ (red and blue curves, respectively) and 
with $\Sigma^{(2)}$ (dashed curve).  
The cluster type dependence (A or B) is negligible.
%[SHOULD WE ALSO SHOW A COMPARISON TO SINGLE-SITE DMFT -> MUCH LESS DAMPING?]
}
\label{current}
\end{figure}

In Fig.~\ref{current}, we plot the current obtained with DCA for $N_c=8\times 8, 16\times 16$  and with $\Sigma^{(2)}$.  
Here, both the cluster types A and B in DCA give almost the same results for these $N_c$.
By comparing the results for $N_c=8\times 8$ and $N_c=16\times 16$, we can confirm that the current converges well 
with respect to $N_c$ up to $tJ\lesssim 5$, which implies that the DCA results can be considered as representative of the thermodynamic limit within this time domain.

One can see that the current shows a coherent Bloch oscillation with frequency $E=4$, but an important question is its damping.  
DMFT predicts a rapid damping of the Bloch oscillation, which is consistent with the previous study
of the dc-field problem for the Hubbard model.\cite{EcksteinWerner2011b}
On the other hand, as we take account of the momentum dependence of the self-energy in DCA or in $\Sigma^{(2)}$, we see a clear difference in the behavior of the oscillation between DCA and $\Sigma^{(2)}$:  
In the DCA case, the current exhibits a longer-lived behavior 
with a beating, i.e., the amplitude of the oscillation oscillates with a longer period, 
while in the $\Sigma^{(2)}$ case, the current is damped monotonically. 
This in itself is physically interesting, and also shows 
that DCA combined with the IPT cluster solver is not equivalent to $\Sigma^{(2)}$, even in the limit of $N_c\to \infty$. DCA can provide more reliable results than $\Sigma^{(2)}$
because the formalism allows one to check the convergence with $N_c$. 
The difference in the results between DCA and $\Sigma^{(2)}$ comes from the fact that DCA imposes a self-consistency condition %with the feedback from the dynamical mean field 
which provides a feedback from the lattice solution  
to the cluster, 
whereas $\Sigma^{(2)}$ does not. If one keeps $N/N_c$ large enough while taking the limit of $N_c\to\infty$ and $N\to\infty$,
the non-trivial effect of this self-consistency may survive.
%In other words, $\lim_{N/N_c\to \infty} \lim_{N_c\to\infty} \mbox{DCA}\neq$

Another advantage of DCA over $\Sigma^{(2)}$ is that there is room for 
improving the cluster solver for DCA,
while it is numerically difficult to extend $\Sigma^{(2)}$ by considering higher-order diagrams for the self-energy
with a large number of $k$ points. In practice, the second-order is the highest for which lattice perturbation theory
can be implement in nonequilibrium.

\section{Summary}
We have formulated the nonequilibrium dynamical cluster theory, which 
enables one to investigate the effects of nonlocal spatial correlations
on nonequilibrium many-body systems by systematically changing the cluster size. We have applied the method to the interaction-quench problem
for the Hubbard model in one and two dimensions, and found a peculiar {\it momentum-dependent relaxation} of quasiparticles in 2D.
This should be experimentally observable by means of time-resolved ARPES measurements, and
such experiments may open an interesting avenue for probing marginal Fermi liquids 
in nonequilibrium.  
We have also applied the method to the Hubbard model driven by a dc electric field, 
and found an enhancement of the Bloch oscillations, compared to the result predicted by DMFT.

Benchmark calculations in 1D revealed a good convergence for local properties, while the accuray of non-local quantites is limited due to our perturbative solution of the impurity problem. Therefore, it will be important to test the cluster approach by also
%Therefore, further improvements of our approach can be achieved by 
using alternative nonequilibrium impurity solvers, such as NCA in the strong-coupling regime,\cite{Eckstein2010} or quantum Monte Carlo (QMC) solvers on smaller clusters in the weak-coupling regime.\cite{WernerOkaMillis2009}
%\dots (I LET YOU FILL IN WHATEVER YOU CAN IMAGINGE TO BE USEFUL ).
While these cluster solvers cannot access large cluster sizes, such as our $N_c=16\times 16$ for the 2D Hubbard model, 
due to the exponentially scaling computational cost, for local and quasilocal quantities such as nearest-neighbor correlation functions the cluster-size dependence can be eliminated even with small clusters.
For example, one may be able to reach $N_c=8$ with the QMC solver or $N_c=4$ with the NCA solver.
The $N_c=8$ cluster can distinguish the nodal [$\bm k=(\pi/2,\pi/2)$] and antinodal [$(\pi,0)$ and $(0,\pi)$] sectors,
where the momentum-dependent relaxation is most evidently observed.
%A first serious benchmarking of the nonequilibrium DMFT and its cluster extension is performed in comparison with the time-dependent DMRG for 1D.
An interesting prospect of the cluster method combined with weak-coupling perturbation theory will be 
the simulation of the real-time dynamics of systems with long-range order (e.g. $d$-wave superconductivity or charge density waves), which would be inaccessible by lattice perturbation theories. 

\acknowledgments 
The calculations were carried out on the UniFr cluster and the Perseus computer cluster of the University of Geneva financed 
by ``Fondation Ernst et Lucie Schmidheiny''. The DMRG simulations employed the ALPS libraries.\cite{Albuquerque2007,Bauer2011} 
N.T. was supported by a 
Grant-in-Aid for Scientific Research on Innovative Areas ``Materials Design through Computics: Complex Correlation
and Non-equilibrium Dynamics'' (Grant No.~25104709), and another 
for Young Scientists (B) (Grant No.~25800192) from MEXT, and SNF Grant No. PP0022-118866. H.A. is supported by a MEXT Grant No.~26247057.  
P.W. acknowledges support from FP7 ERC Starting Grant No. 278023.

\bibliographystyle{apsrev}
\bibliography{noneq-dca}

\end{document}